\documentclass[a4paper]{jpconf}
\usepackage[T1]{fontenc}
\usepackage[latin1]{inputenc}
\usepackage{amsmath}

\begin{document}

\title{Observational Constraints on Transverse Gravity: a Generalization
of Unimodular Gravity.}

\author{J.J. Lopez-Villarejo}

\address{Departamento de Fisica Teorica.\\
Universidad Autonoma de Madrid\\
Campus de Cantoblanco, 28049, Madrid (Spain).}

\ead{jj.lopezvillarejo@uam.es}

\begin{abstract}
We explore the hypothesis that the set of symmetries enjoyed by the
theory that describes gravity is not the full group of diffeomorphisms
(\emph{Diff(M)}), as in General Relativity, but a maximal subgroup
of it (\emph{TransverseDiff(M)}), with its elements having a jacobian
equal to unity; at the infinitesimal level, the parameter describing
the coordinate change $x^{\mu}\rightarrow x^{\mu}+\xi^{\mu}(x)$ is
transverse, i.e., $\partial_{\mu}\xi^{\mu}=0$. Incidentally, this
is the smaller symmetry one needs to propagate consistently a graviton,
which is a great theoretical motivation for considering these theories.
Also, the determinant of the metric, $g$, behaves as a {}``transverse
scalar'', so that these theories can be seen as a generalization
of the better-known unimodular gravity. We present our results on
the observational constraints on transverse gravity, in close relation
with the claim of equivalence with general scalar-tensor theory. We
also comment on the structure of the divergences of the quantum theory
to the one-loop order. 
\end{abstract}

\section{Introduction \& Formalism}

The gravitational theory of General Relativity presents as a remarkable
feature the property of (active) diffeomorphism gauge invariance.
The Einstein-Hilbert action\begin{equation}
S_{EH}=-\frac{1}{2\kappa}\int d^{4}x\sqrt{g}R\label{eq:einstein-hilbert action}\end{equation}
is symmetric%
\footnote{Unlike internal gauge transformations which do not involve the coordinates
(e.g., U(1) of electromagnetism), this gauge invariance is not attained
at the level of the Lagrangian, but only at the level of the action.%
} under the {}``Diff'' transformation $g_{\mu\nu}(x)\rightarrow g_{\mu\nu}(x)+\nabla_{\mu}\xi_{\nu}(x)+\nabla_{\nu}\xi_{\mu}(x)$,
which represents the active version of the infinitesimal coordinate
transformation $x^{\mu}\rightarrow x^{\mu}+\xi^{\mu}(x)$. 

When attempting to quantize this theory in a perturbative expansion,
one can take the ansatz $g_{\mu\nu}=\widetilde{g}_{\mu\nu}+h_{\mu\nu}$,
where $\widetilde{g}_{\mu\nu}$ is a classical background that fulfills
the classical equations of motion (e.o.m), and $h_{\mu\nu}$ is the
quantum perturbation. In this way, one gets\begin{equation}
S=-\frac{1}{2\kappa}\int d^{4}x[const+o(h_{\mu\nu}^{2})+o(h_{\mu\nu}^{3})+...]\label{eq:}\end{equation}
where the quadratic part in $h_{\mu\nu}$ constitutes the free-theory.
In the case of a flat space, with a Minkowski classical background,
the free-theory is given by the Fierz-Pauli Lagrangian:\begin{equation}
\mathcal{L}_{FP}=\mathcal{L}^{I}+\mathcal{L}^{II}+a\mathcal{L}^{III}+b\ \mathcal{L}^{IV}\,\,\,\, with\, b=\left(1-2a+3a^{2}\right)/2\label{eq:Fierz-Pauli}\end{equation}
where $\mathcal{L}^{I}\equiv\frac{1}{4}\partial_{\mu}h^{\nu\rho}\partial^{\mu}h_{\nu\rho}$,
$\mathcal{L}^{II}=-\frac{1}{2}\partial_{\mu}h^{\mu\rho}\partial_{\nu}h_{\rho}^{\nu}$,
$\mathcal{L}^{III}=\frac{1}{2}\partial^{\mu}h\partial^{\rho}h_{\mu\rho}$,
$\mathcal{L}^{IV}=-\frac{1}{4}\partial_{\mu}h\partial^{\mu}h$. Note
that we have allowed for field redefinitions $h_{\mu\nu}\rightarrow\phi h_{\mu\nu}$
with respect to the usual presentation of the Fierz-Pauli Lagrangian,
when $a=b=1$. The corresponding action is symmetric under the {}``linear
Diff'' transformation\begin{equation}
h_{\mu\nu}\rightarrow h_{\mu\nu}+\partial_{\mu}\xi_{\nu}+\partial_{\nu}\xi_{\mu}\label{eq:}\end{equation}
and represents the action for a free massless spin-two particle (graviton).
Here, gauge invariance is essential to eliminate the unphysical propagating
modes, much like in electromagnetism, where the bad-behaved longitudinal
mode is made a gauge artifact through the symmetry $A_{\mu}\rightarrow A_{\mu}+\partial_{\mu}\alpha$. 

The authors of \cite{vanderBij:1981ym}  undertook the task of finding
out the most general conditions for a consistent description of the
free massless spin-two particle. This involved considerations regarding
the little group ISO(2) of the masless spin-two particle and the necessity
of triviality of the generators of translations. Their conclusion
is that, actually, one only needs the somehow smaller gauge symmetry\begin{equation}
h_{\mu\nu}\rightarrow h_{\mu\nu}+\partial_{\mu}\xi_{\nu}+\partial_{\nu}\xi_{\mu},\,\,\,\,\, with\,\,\partial_{\mu}\xi^{\mu}=0,\label{eq:Linear Tdiff transformation}\end{equation}
which we will denote as the {}``Linear TDiff'' transformation, to
indicate the transverse condition on the vector parameter. Hence,
one can think of alternative theories of gravity to General Relativity,
based on a slightly different symmetry principle, which however contain
the graviton in the linear regime as well. From this observation,
there are basically two ways to proceed forward.

The first route was already explored in the aforementioned seminal
paper \cite{vanderBij:1981ym}. One should note that the transversality
condition in \eqref{eq:Linear Tdiff transformation} is directly affecting
the transformation rule of the trace of $h_{\mu\nu}$: $h_{\mu}^{\mu}\rightarrow h_{\mu}^{\mu}+\partial_{\mu}\xi^{\mu}$,
such that this quantity is an invariant. Therefore, one can decide
to restrict the value of the trace in the theory to have a fixed value,
tipically $h_{\mu}^{\mu}=0$. In the full non-linear regime, this
naturally translates into the condition $g\equiv-det(g_{\mu\nu})=1$.
This approach has been named \emph{unimodular gravity. }

In this work we are concerned with the second way to proceed. One
can ask himself about the most general theory involving an unconstrained
second-rank tensor $h_{\mu\nu}$ and which is compatible with the
TDiff symmetry. Since we will be using less symmetry than the standard
Diff case, we expect to get a \emph{more general} action also in the
full non-linear regime:\begin{equation}
S_{TDiff}=S_{TDiff}[g_{\mu\nu},\psi]=S_{Diff}[g_{\mu\nu},\psi]+S'[g_{\mu\nu}(x),\psi].\label{eq:}\end{equation}
We wondered whether this fact could be useful in addressing such problems
as the dark energy of dark matter puzzles.

The concrete answer in the linear regime was found in \cite{AlvarezVerdagerGarrigaBlas:2006uu}.
In general, one gets an extra scalar-like mode propagating as well
as the two tensor modes corresponding to the massless graviton ---
the only exception is when the TDiff symmetry is enhanced with an
additional Weyl conformal symmetry, in which case the unimodular approach
is recovered. Specifically:\begin{equation}
\mathcal{L}_{L-TDiff}=\mathcal{L}^{I}+\mathcal{L}^{II}+a\mathcal{L}^{III}+b\ \mathcal{L}^{IV}\,\,\,\, with\, a,\, b\,\, arbitrary\label{eq:TDiff linear lagrangian}\end{equation}
where the form of the different pieces in the Lagrangian is given
below \eqref{eq:Fierz-Pauli}. It should be stressed that this TDiff
Langrangian, as well as the following ones, are not covariant with
respect to general coordinate transformations%
\footnote{In passing, one could render the theory covariant by introducing extra
objects in it.%
}, so that one has to formulate them in a specific set of coordinates.
This is the reason why we are not talking of a \emph{proper} additional
scalar mode. 

In the non-linear regime, the symmetry \eqref{eq:Linear Tdiff transformation}
\emph{naturally} generalizes%
\footnote{This is, however, not a unique answer. The other most {}``sensible''
generalization, where $\xi^{\mu}$ is restricted with $\nabla_{\mu}\xi^{\mu}=0$
(and $\sqrt{g}\rightarrow\sqrt{g}$) is doomed since one cannot accommodate
a kinetic term for $g$: $\frac{1}{2}f_{k}(g)g^{\mu\nu}\partial_{\mu}g\partial_{\nu}g$.%
} to\begin{equation}
g_{\mu\nu}\rightarrow g_{\mu\nu}+\nabla_{\mu}\xi_{\nu}+\nabla_{\nu}\xi_{\mu},\,\,\,\,\, with\,\,\partial_{\mu}\xi^{\mu}=0\label{eq:}\end{equation}
\begin{equation}
\Longrightarrow\,\,\,\,\sqrt{g}\rightarrow\sqrt{g}+\xi^{\mu}\partial_{\mu}\sqrt{g}\label{eq:}\end{equation}
Hence, the determinant of the metric acts as a scalar under this restricted
transformation (it is not, however, a scalar under coordinate transformations).
What this entails is that one can have arbitrary powers of $g$ appearing
everywhere in the action of the full TDiff theory:\begin{eqnarray}
S_{TDiff} & = & -\frac{1}{2\kappa^{2}}\int_{\mathcal{M}}d^{4}x\sqrt{g}\left(f(g)R+2f_{\lambda}(g)\Lambda+\frac{1}{2}f_{k}(g)g^{\mu\nu}\partial_{\mu}g\partial_{\nu}g\right)\label{eq:Full TDiff action}\\
 &  & +\int_{\mathcal{M}}d^{4}x\sqrt{g}\mathcal{L}_{Matter}[\psi,g_{\mu\nu};\, g],\nonumber \end{eqnarray}
formulated in a specific set of coordinates.

\section{Results}

The first question that was examined was the ultraviolet behaviour
of these theories. Less symmetry translates in this case into more
divergences, as compared with the template of General Relativity.
Specifically, a calculation employing the background field method
was undertaken in order to extract the 1-loop order divergences \cite{AlvarezFaedoVillarejoUV:2008zw}.
We made use of a correspondence between TDiff theories and scalar-tensor
theories in the so-called {}``unitary gauge'', which helped simplify
the computations. To 1-loop order, General Relativity is UV (on-shell)
finite \cite{'tHooft:1974bx}. These theories are not, due to extra
terms arising from the additional $g$-mode; in fact, the result is
very similar to a scalar-tensor theory with the scalar replaced by
$g$. The only exceptions are two already-mentioned limiting cases:
General Relativity and the presence of an additional conformal Weyl
symmetry (unimodular case).

Then, the observational signatures of these kind of theories were
inspected \cite{AlvarezFaedoVillarejoobser:2009ga}. A general remark
is that an almost exact correspondence can be found between these
theories and scalar-tensor theories where the $g$ is promoted to
a true scalar field $\phi$, and a lagrangian multiplier is added
to the action to force the constraint $\phi=g$ (the precise form
of this correspondence remains to be stablished {[}in preparation]).
In practice, we can thus apply all existing experimental bounds on
scalar-tensor gravity to constrain these theories. One should note
that, in general, the presence of $g$ {}``alone'' in the matter
action will translate into a \emph{non-metric coupling} of the scalar
$\phi$, when viewed in terms of the scalar-tensor language; hence,
we will get wild violations of the (Weak) Equivalence Principle. A
mechanism in the spirit of the metric postulate \cite{Will:2005va},
or the one usually advocated in string theory \cite{Damour:2001fn}
will be generically necessary to render the theory compatible with
observations.

\section{Conclusions}

\begin{itemize}
\item TDiff theories are well motivated alternative theories of gravity
from a quantum perspective.
\item To the 1-loop level, these theories present UV divergences --- while
General Relativity is UV finite ---, with only one significant exception
that corresponds, arguably, to Unimodular Gravity.
\item The experimental bounds on scalar-tensor theories of gravity are generically
applicable to these theories. A mechanism to avoid {}``excessive''
violations of the (Weak) Equivalence Principle is thus necessary to
render the theory compatible with experiments.
\end{itemize}


\begin{thebibliography}{7}  

\bibitem{vanderBij:1981ym}   J.~J.~van der Bij, H.~van Dam and Y.~J.~Ng,   ``The Exchange Of Massless Spin Two Particles,''   Physica {\bf 116A}, 307 (1982). 

\bibitem{AlvarezVerdagerGarrigaBlas:2006uu}   E.~Alvarez, D.~Blas, J.~Garriga and E.~Verdaguer,   ``Transverse Fierz-Pauli symmetry,''   Nucl.\ Phys.\  B {\bf 756} (2006) 148   [arXiv:hep-th/0606019].

\bibitem{AlvarezFaedoVillarejoUV:2008zw}   E.~Alvarez, A.~F.~Faedo and J.~J.~Lopez-Villarejo,   ``Ultraviolet behavior of transverse gravity,''   JHEP {\bf 0810} (2008) 023   [arXiv:0807.1293 [hep-th]].  

\bibitem{'tHooft:1974bx}   G.~'t Hooft and M.~J.~G.~Veltman,   ``One loop divergencies in the theory of gravitation,''   Annales Poincare Phys.\ Theor.\  A {\bf 20} (1974) 69.  

\bibitem{AlvarezFaedoVillarejoobser:2009ga}   E.~Alvarez, A.~F.~Faedo and J.~J.~Lopez-Villarejo,   ``Observational constraints on transverse gravity,''   JCAP {\bf 0907} (2009) 002   [arXiv:0904.3298 [hep-th]].  

\bibitem{Will:2005va}   C.~M.~Will,   ``The confrontation between general relativity and experiment,''   Living Rev.\ Rel.\  {\bf 9}, 3 (2005)   [arXiv:gr-qc/0510072].   

\bibitem{Damour:2001fn}   T.~Damour,   ``Questioning the equivalence principle,''   arXiv:gr-qc/0109063.  
  

\end{thebibliography}
\end{document}